\input harvmac
\input epsf
\def\aPK{a_{\psi K_S}}
\def\aPP{a_{\pi\pi}}
\def\epsK{\varepsilon_K}
\def\H{{\cal H}}
\def\ra{\rightarrow}
\def\td{\theta_d}
\def\al{\alpha}
\def\be{\beta}
\def\mod{{\rm mod}}

\def\ra{\rightarrow}

\def\gsim{{~\raise.15em\hbox{$>$}\kern-.85em
          \lower.35em\hbox{$\sim$}~}}
\def\lsim{{~\raise.15em\hbox{$<$}\kern-.85em
          \lower.35em\hbox{$\sim$}~}}
\def\SM{Standard Model}
\def\np{new physics}
\def\Np{New physics}
\def\UT{unitarity triangle}
\def\bb{B-\bar B}
\def\YGTitle#1#2{
\nopagenumbers\abstractfont\hsize=\hstitle\rightline{#1}%
\vskip .6in\centerline{\titlefont #2}\abstractfont\vskip .3in\pageno=0}
\YGTitle{\vbox{
   \hbox{SLAC-PUB-7450}
   \hbox{WIS-97/10/Apr-PH}
   \hbox{hep-ph/9704287}}}
{\vbox{\centerline{A Model Independent Construction of the}
\vbox{\vskip 0.truecm}
\vbox{\centerline{Unitarity Triangle}}}}
\smallskip
\centerline{Yuval Grossman$^a$, Yosef Nir$^b$ and Mihir P. Worah$^a$}
\smallskip
\bigskip
\centerline{\it $^a$Stanford Linear Accelerator Center,
 Stanford University, Stanford, CA 94309, USA}
\centerline{\it $^b$Department of Particle Physics,
 Weizmann Institute of Science, Rehovot 76100, Israel}
\smallskip
\bigskip
\baselineskip 18pt
\noindent

In a large class of models, the only significant \np\
effect on the CP asymmetries in $B\ra\psi K_S$ and $B\ra\pi\pi$ decays
is a new contribution to the $\bb$ mixing amplitude.
This allows a model independent construction of the CKM Unitarity
Triangle (up to hadronic uncertainties). Furthermore, the contributions to
the mixing from the Standard Model and from the \np\
can be disentangled. A serious obstacle to this analysis
is an eightfold discrete ambiguity in solving for the angles
of the triangle. Several ways to reduce the ambiguity either by
making further measurements, or by making further assumptions
about the nature of the \np\ are described.
\bigskip

\baselineskip 18pt
\leftskip=0cm\rightskip=0cm

\Date{}

\newsec{The Basic Assumptions and Results}

The first two CP asymmetries to be measured in a $B$ factory
are likely to be
\ref\NiQu{For a review see
Y. Nir and H.R. Quinn, Ann. Rev. Nucl. Part. Sci. 42 (1992) 211.}%
\eqn\firstaCPb{
{\Gamma(B^0_{\rm phys}(t)\ra\psi K_S)-
\Gamma(\bar B^0_{\rm phys}(t)\ra\psi K_S)\over
\Gamma(B^0_{\rm phys}(t)\ra\psi K_S)+
\Gamma(\bar B^0_{\rm phys}(t)\ra\psi K_S)}= -\aPK\sin(\Delta m_B t),}
\eqn\firstaCPa{
{\Gamma(B^0_{\rm phys}(t)\ra\pi\pi)-
\Gamma(\bar B^0_{\rm phys}(t)\ra\pi\pi)\over
\Gamma(B^0_{\rm phys}(t)\ra\pi\pi)+
\Gamma(\bar B^0_{\rm phys}(t)\ra\pi\pi)}= -\aPP\sin(\Delta m_B t).}
%
In addition, the $B$ factory will improve our knowledge of the $B-\bar B$
mixing parameter, $x_d\equiv{\Delta m_B\over\Gamma_{B}}$, and of the
charmless semileptonic branching ratio of the $B$ mesons.

Within the \SM, these four measurements are useful in
constraining the \UT. The asymmetries \firstaCPb\ and \firstaCPa\
measure angles of the \UT:
\eqn\SMaPK{\aPK=\ \sin2\beta,}
\eqn\SMaPP{\aPP=\ \sin2\alpha,}
where
\eqn\SMangles{\alpha\equiv\
\arg\left[-{V_{td}V_{tb}^*\over V_{ud}V_{ub}^*}\right],\ \ \
\beta\equiv\
\arg\left[-{V_{cd}V_{cb}^*\over V_{td}V_{tb}^*}\right].}
In \SMaPK\ we have taken into account the fact that the final
state is CP-odd.
In \SMaPP\ we have ignored possible penguin contamination which can,
in principle, be eliminated by isospin analysis
\ref\GrLo{M. Gronau and D. London, Phys. Rev. Lett. 65 (1990) 3381.}.
The measurement of $x_d$ determines one side of the \UT\ ($R_t$):
\eqn\SMxd{x_d=C_t R_t^2,}
where
\eqn\defRt{R_t\equiv\left|{V_{tb}^*V_{td}\over V_{cb}^*V_{cd}}\right|,}
and $C_t=\tau_b{G_F^2\over6\pi^2}\eta_B m_B(B_Bf_B^2)m_t^2
f_2(m_t^2/m_W^2)|V_{cb}^*V_{cd}|^2$ (for definitions and notations see
\NiQu).
The present values are $x_d= 0.73 \pm 0.05$ and
$C_t \sim 0.4 - 0.8$ for $\sqrt{B_B}f_B = 140-200$ MeV
\ref\pdg{R.M. Barnett {\it et al.} [Particle Data Group], 
Phys. Rev. D54 (1996) 1.}.
Measurements of various inclusive and exclusive $b\ra u\ell\nu$ processes
will determine 
(up to uncertainties arising from various hadronic models) 
the length of
the other side of the \UT\ ($R_u$):
\eqn\SMbu{{\Gamma(b\ra u\ell\nu)\over\Gamma(b\ra c\ell\nu)}=
{1\over F_{{\rm ps}}}\left|{V_{cd}\over V_{ud}}\right|^2 R_u^2,}
where
\eqn\defRb{R_u\equiv\left|{V_{ub}^*V_{ud}\over V_{cb}^*V_{cd}}\right|}
and $F_{\rm ps}\approx0.5$ is a phase space factor.
The present value for $R_u$ ranges from 0.27 to 0.45 depending on the
hadronic model used to relate the measurement at the
endpoint region, or of some exclusive mode, to the total
$b \ra u$ inclusive rate \pdg.

In the presence of \np\ it is quite possible that the \SM\
predictions \SMaPK, \SMaPP\ and \SMxd\ are violated. The most likely
reason is a new, significant contribution to $B-\bar B$ mixing that
carries a CP violating phase different from the \SM\ one.
Other factors that could affect the construction of the \UT\
from these four measurements are unlikely to be significant
\nref\NiSi{Y. Nir and D. Silverman, Nucl. Phys. B345 (1990) 301.}%
\nref\DLN{C.O. Dib, D. London and Y. Nir,
 Int. J. Mod. Phys. A6 (1991) 1253.}%
\refs{\NiSi-\DLN}:
\item{a.} The $\bar b\ra\bar cc\bar s$ and $\bar b\ra\bar uu\bar d$
decays for $\aPK$ and $\aPP$ respectively,
as well as the semileptonic $B$ decays for $R_u$, are mediated by
\SM\ tree level diagrams. In most extensions of the Standard Model
there is no decay mechanism that could significantly compete with
these contributions. (For exceptions, 
which could affect
the $\bar b\ra\bar uu\bar d$ decay, see
\ref\GrWo{Y. Grossman and M. Worah, Phys. Lett. B395 (1997) 241.}.)
\item{b.} \Np\ could contribute significantly to $K-\bar K$
mixing. However, the small value of $\epsK$ forbids large deviations
from the \SM\ phase of the mixing amplitude. 
%
\item{c.} Unitarity of the three generation CKM matrix is maintained
if there are no quarks beyond the three generations of the Standard
Model. Even in models with an extended quark sector the effect on
$B-\bar B$ mixing is always larger than the violation
of CKM unitarity. 

Our analysis below applies to models where the above three conditions are
not significantly violated. Under these circumstances the relevant \np\
effects can be described by two new parameters, $r_d$ and $\theta_d$
\nref\SoWo{J.M. Soares and L. Wolfenstein, Phys. Rev. D47 (1993) 1021.}%
\nref\DDO{N.G. Deshpande, B. Dutta and S. Oh,
 Phys. Rev. Lett. 77 (1996) 4499.}%
\nref\SiWo{J.P. Silva and L. Wolfenstein, hep-ph/9610208.}%
\nref\CKLN{A.G. Cohen, D.B. Kaplan, F. Lepeintre and A.E. Nelson,
 Phys. Rev. Lett. 78 (1997) 2300.}%
\refs{\SoWo-\CKLN},
defined by
\eqn\thetad{\left(r_d e^{i\theta_d}\right)^2\equiv{
\vev{B^0|\H^{\rm full}_{\rm eff}|\bar B^0}\over
\vev{B^0|\H^{\rm SM}_{\rm eff}|\bar B^0}},}
where $\H^{\rm full}_{\rm eff}$ is the effective Hamiltonian
including both \SM\ and \np\ contributions, and
$\H^{\rm SM}_{\rm eff}$ only includes the \SM\ box diagrams.
In particular, with this definition, the modification of the two CP
asymmetries in \SMaPK\ and \SMaPP\ depends on a {\it single} new
parameter, the phase $\td$:
\eqn\NPaPK{\aPK=\ \sin(2\beta+2\theta_d),}
\eqn\NPaPP{\aPP=\ \sin(2\alpha-2\theta_d),}
while the modification of the $B-\bar B$ mixing parameter $x_d$
in \SMxd\ is given by the magnitude rescaling parameter, $r_d$:
\eqn\NPxd{x_d=C_t R_t^2 r_d^2.}
Furthermore, since the determination of $R_u$ from the
semileptonic $B$ decays is not affected by the \np, and
since the \UT\ remains valid, we have the following
relations between the length of its sides and its angles:
\eqn\Rbab{R_u={\sin\beta\over\sin\alpha},}
\eqn\Rtab{R_t={\sin\gamma\over\sin\alpha},}
where
\eqn\defgamon{
\gamma\equiv\arg\left[-{V_{ud}V_{ub}^*\over V_{cd}V_{cb}^*}\right].}
When $\al$, $\be$ and $\gamma$
are defined to lie in the $\{0,2\pi\}$ range, they satisfy
\eqn\defgam{\alpha+\beta+\gamma=\pi\ {\rm or}\ 5\pi.}

The four measured quantities $\aPK$, $\aPP$, $x_d$ and $R_u$
can be used to achieve the following \SoWo:
\item{(i)} Fully reconstruct the unitarity triangle and,
in particular, find $\alpha$, $\beta$ and $R_t$;
\item{(ii)} Find the magnitude and phase of the \np\ contribution
to $B-\bar B$ mixing, namely determine $r_d$ and $\theta_d$.

It is straightforward to show that the above tasks are possible.
Eqs. \NPaPK, \NPaPP\ and \Rbab\ give three equations for three unknowns,
$\alpha$, $\beta$ and $\theta_d$. Once $\alpha$ and $\beta$ are known,
$\gamma$ can be extracted from \defgam, $R_t$ can then be deduced from
\Rtab, and finally $r_d$ is found from \NPxd.

In practice, however, it is quite likely that the combination of
experimental and theoretical uncertainties (particularly in
the $x_d$ and $R_u$ constraints) and discrete ambiguities
will limit the usefulness of the above method rather significantly.
In the next section we discuss the discrete ambiguities that arise in
this calculation. We then describe how to determine the
parameters, both in the $\rho-\eta$ plane (section 3), and in the
$\sin 2\al-\sin 2\be$ plane (section 4). We mention ways to
resolve some of the ambiguities in the concluding section.

\newsec{Discrete Ambiguities}

A major obstacle in carrying out the above program will be the discrete
ambiguities in determining $\gamma$. We now describe these ambiguities.

A physically meaningful range for an angle is $2\pi$.
We choose this range to be $\{0,2\pi\}$.
Measurement of any single asymmetry,
$\sin2\phi$, determines the corresponding angle only up to a fourfold
ambiguity: $\phi$, $\pi/2-\phi$, $\pi+\phi$ and $3\pi/2-\phi$ (mod $2\pi$).
Specifically, let us denote by $\bar\alpha$ and $\bar\beta$
some solution of the equations
\eqn\somesol{\aPK=\sin2\bar\beta,\ \ \ \aPP=\sin2\bar\alpha.}
Thus, measurements of the two asymmetries leads to a sixteenfold
ambiguity in the values of the $\{\bar\al,\bar\be\}$ pair.
However, since $\bar\al=\al-\theta_d$ and $\bar\be=\be+\theta_d$, and
unitarity is not violated, $\gamma$ still satisfies the condition
\eqn\NPsum{\bar\al+\bar\be+\gamma=\pi\ (\mod\ 2\pi).}
Then, the sixteen possibilities for $\gamma$ are divided into
two groups of eight that are related by the combined operation
$\bar\al \ra \bar\al+\pi$ and $\bar\be \ra \bar\be+\pi$.
This, in turn shifts the value of $\gamma$ by $2\pi$.
However, since $\gamma$ is only defined modulo $2\pi$,
the ambiguity in $\gamma$ is reduced to eightfold.
We emphasize that this reduction of the ambiguity
depends only on the definition of $\gamma$. Defining
\eqn\defphipm{\phi_\pm=\bar\alpha\pm\bar\beta,}
the eight possible solutions for $\gamma$ are
\eqn\MIgamma{\gamma=\pm\phi_+,\ \pi\pm\phi_+,\ \pi/2\pm\phi_-,\
3\pi/2\pm\phi_-\ (\mod\ 2\pi).}
Note that the eight solutions come in pairs of $\pm\gamma$.
This in turn implies that the ambiguity on $R_t$ is only fourfold.

In any model where the three angles $\bar\al$, $\bar\be$, and $\gamma$ form a
triangle, the ambiguity is further reduced
\ref\NiQu{Y. Nir and H.R. Quinn, Phys. Rev. D42 (1990) 1473.}:
the requirement that the angles are either all in the range
$\{0,\pi\}$ or all in the range $\{\pi,2\pi\}$ reduces the ambiguity
in $\gamma$ to fourfold. It is enough to know the signs of $\aPK$ and
$\aPP$ to carry out this step. Finally, within the \SM, the bound
$0<\beta<\pi/4$ (obtained from the sign of $\epsK$ and from
$R_u < 1/\sqrt{2}$) reduces the ambiguity in $\gamma$ to twofold.

When we allow for the possibility of \np\ effects in the mixing,
knowing the signs of $\aPK$ and $\aPP$ does not lead to further reduction
in the ambiguity, which remains eightfold.
The three angles $\bar\alpha$, $\bar\beta$ and
$\gamma$ are not angles that define a triangle and therefore further
constraints cannot be imposed. It is possible, for example, that
both $\gamma$ and $\bar\beta$ lie in the range $\{\pi/2,\pi\}$.
Further the sign of $\epsK$ may not be related to the sign of $\eta$.

The following example will make the situation clear. Take
\eqn\examplea{\aPP=1/2,\ \ \ \aPK=\sqrt3/2.}
Then, we could have
\eqn\apria{
\bar\alpha={\pi\over12},{5\pi\over12},{13\pi\over12},{17\pi\over12},\ \
\bar\beta={\pi\over6},{\pi\over3},{7\pi\over6},{4\pi\over3}.}
The eight solutions for $\gamma$ are
\eqn\solveaMI{\gamma={\pi \over 4},\ {5\pi\over 12},\ {7\pi\over 12},\
{3\pi\over 4},\
{5\pi\over 4},\ {17\pi\over 12},\ {19\pi\over 12},\ {7\pi\over 4}.}
%
If $\bar\alpha,\bar\beta,\gamma$
define a triangle, then only four solutions are allowed:
\eqn\solvea{(\bar\alpha,\bar\beta,\gamma)=
\left({\pi\over12},{\pi\over6},{3\pi\over4}\right),
\left({\pi\over12},{\pi\over3},{7\pi\over12}\right),
\left({5\pi\over12},{\pi\over6},{5\pi\over12}\right),
\left({5\pi\over12},{\pi\over3},{\pi\over4}\right).}
Assuming $0<\bar\beta<\pi/4$ as in the \SM\ leaves only the first
two choices.

In various specific cases, the discrete ambiguity is smaller. If the two
asymmetries are equal in magnitude,
there is only a sixfold ambiguity:
\eqn\MIequal{\eqalign{\aPP=\aPK\ \Longrightarrow&\ \gamma=
\pm2\bar\beta,\ \pi\pm2\bar\beta,\ \pi/2,\ 3\pi/2\ (\mod\ 2\pi),\cr
\aPP=-\aPK\ \Longrightarrow&\ \gamma=0,\ \pi,\ \pi/2\pm2\bar\beta,\
3\pi/2\pm2\bar\beta\ (\mod\ 2\pi).\cr}}
If one of the asymmetries is maximal, there is a fourfold ambiguity, e.g.
\eqn\MIma{\eqalign{\aPP=+1\ \Longrightarrow&\ \gamma=
\pm(\pi/4+\bar\beta),\ \ \pm(3\pi/4-\bar\beta)\ \ (\mod\ 2\pi),\cr
\aPP=-1\ \Longrightarrow&\ \gamma=\pm(\pi/4-\bar\beta),\ \
\pm(3\pi/4+\bar\beta)\ \ (\mod\ 2\pi).\cr}}
If both asymmetries are maximal, the ambiguity is twofold.
If the two asymmetries vanish, there is only a fourfold ambiguity:
\eqn\MICP{\aPP=\aPK=0\ \Longrightarrow\
\gamma=0,\ \pi/2,\ \pi,\ 3\pi/2.}
This is an interesting case, because it is predicted by models
with approximate CP symmetry (e.g. in some supersymmetric models
\ref\GNR{For a review, see Y. Grossman, Y. Nir and R. Rattazzi,
hep-ph/9701231.}). Only two of the solutions ($0,\pi$) correspond to the
CP symmetric case while in the other two ($\pi/2,3\pi/2$), the zero
asymmetries are accidental.

So far we have ignored the penguin contamination in $\aPP$.
The isospin analysis eliminates the penguin contamination only up to
a four fold ambiguity \GrLo.
Therefore, if the isospin analysis is needed, the ambiguities are
increased.

In addition, for each value of $\gamma$ there are two
possibilities for $\td$ related by $\td \ra \td+\pi$.
As long as the \np\ is such that the $\Delta b=2$ operator
that contributes to $B-\bar B$ mixing can be separated into two
$\Delta b=1$ operators the $\td \ra \td+\pi$ ambiguity is physical.
Otherwise, it is not physical.

\newsec{The $\rho-\eta$ Plane}

The key point in the extraction of the CKM parameters is that the
angle $\theta_d$ cancels in the following sum:
\eqn\indsum{2(\alpha+\beta)=\arcsin(\aPK)+\arcsin(\aPP).}
In other words, the angle $\gamma$ can be determined (up to the
discrete ambiguities discussed above). In the $\rho-\eta$ plane, a value
for $\gamma$ gives a ray from the origin, while a value for $R_u$ gives
a circle that is centered in the origin. The intersection point
of the line and the circle gives $(\rho,\eta)$ of the unitarity
triangle and determines it completely.

A graphical way to carry out these calculations in the $\rho-\eta$ plane
is the following (see Figure 1) \CKLN. One draws the four curves that
correspond to eqs. \SMaPK, \SMaPP, \SMxd\ and \SMbu\ (even though only
the latter is valid!). The next step is to draw the ray from the origin
that passes through the intersection point of the $\beta$-ray and the
$\alpha$-circle: this is the {\it correct} $\gamma$-ray (see the dashed
line in Figure 1). The intersection point of the $\gamma$-ray and the
$R_u$-circle gives the {\it correct} vertex of the \UT, $(\rho,\eta)$,
namely
\eqn\trueUT{\eqalign{\tan\beta=&\ {\eta\over1-\rho},\cr
R_t^2=&\ \eta^2+(1-\rho)^2.\cr}}

The information about the \np\ contribution to $B-\bar B$ mixing
is found from the intersection point of the $\beta$-ray and the
$x_d$-circle, ($\rho',\eta'$), namely
\eqn\findtdRd{\eqalign{
\theta_d=&\ \arctan{\eta'\over1-\rho'}-\arctan{\eta\over1-\rho},\cr
r_d^2=&\ {\eta'^2+(1-\rho')^2\over \eta^2+(1-\rho)^2}.\cr}}

\newsec{The $\sin2\alpha-\sin2\beta$ Plane}

A presentation of the various constraints in the $\sin2\alpha-\sin2\beta$
plane
\nref\NiSa{Y. Nir and U. Sarid, Phys. Rev. D47 (1993) 2818.}%
\nref\GrNibsg{Y. Grossman and Y. Nir, Phys. Lett. B313 (1993) 126.}%
\refs{\SoWo,\NiSa,\GrNibsg}\ is useful because the two angles are usually
correlated
\ref\DDGN{C.O. Dib, I. Dunietz, F.J. Gilman and Y. Nir,
 Phys. Rev. D41 (1990) 1522.}.
The model independent analysis is demonstrated in Figure 2. The
$R_u$ constraint gives an eight-shaped curve on which the physical
values have to lie.
The various solutions for eq. \indsum\ fall on two ellipses, the
intersections of which with the $R_u$ curve determine the allowed
values of $\sin2\al$ and $\sin2\be$.
Note that these ellipses cross the eight-shaped curve in sixteen points
but, as argued above, only eight of these points are true solutions.
The inconsistent intersection points can be found by noting that the
slopes of the ellipse at the consistent points should be
$(\cos2\alpha,-\cos2\beta)$. The eight correct solutions are
denoted by the filled circles in Figure 2.

In the above, we showed how to use measured values of the CP asymmetries
$\aPK$ and $\aPP$ to find the allowed values for $\alpha$ and $\beta$.
The presentation in the $\sin2\alpha-\sin2\beta$ plane is also useful
for the opposite situation. Some models predict specific values for
$\alpha$ and $\beta$. (Such predictions can arise naturally from
horizontal
symmetries.) On the other hand, the models often allow new contributions
to $B-\bar B$ mixing of unknown magnitude and phase. In this case, the
predicted value of $(\sin2\alpha,\sin2\beta)$ is just a point in the
plane, and the ellipse \indsum\ actually gives the allowed (and
correlated) values of $(\aPP,\aPK)$. Such an analysis was carried out in
ref. \ref\BHR{R. Barbieri, L.J. Hall and A. Romanino, hep-ph/9702315.}.

More generally, even in models that make no specific predictions
for CKM parameters, we usually have some constraints on the allowed
range for $\alpha$ and $\beta$. For example, in this work we assume
the validity of the limits on $R_u$ from charmless semileptonic
$B$ decays which constrains the ratio $\sin\beta/\sin\alpha$ through
\Rbab. Note, however, that this constraint by itself cannot exclude
any region in the $\aPP-\aPK$ plane. The reason is the following.
For any value of $R_u$, neither $\alpha$ nor $\td$ are constrained.
(The angle $\beta$ is constrained for any $R_u<1$ and certainly by
the present range, $0.27<R_u<0.45$.)
Then any value of $\aPK$ can be accommodated by an appropriate
choice of $\td$ and any value of $\aPP$ can be fitted by further
choosing an appropriate $\alpha$. Obviously, to get predictions
for the CP asymmetries beyond the Standard Model, one has to make
some assumptions that go beyond our generic analysis.

For example, consider models where $\epsK$ is dominated
by the \SM\ box diagrams (while $B-\bar B$ mixing is not). Then, we know
that $0<\gamma<\pi$. This already excludes part of the allowed
range. In particular, $(\aPP,\aPK)=(1,-1)$ or $(-1,1)$ requires
$\gamma=0$ or $\pi$, and is therefore excluded in this class of
models. More generally, in any class of models where $\sin^2\gamma$
cannot assume any value between zero and one,
some regions in the $\aPP-\aPK$ plane are excluded.

\newsec{Final Comments}

We argued that the most likely effect of \np\ on CP asymmetries
in neutral $B$ decays into CP eigenstates will be
a significant contribution to the mixing. This is because we have
concentrated on decays that are allowed at tree level in the \SM. Thus
the \np\ effects on
the decay amplitudes and on CKM unitarity can be neglected in a large
class of models%
\foot{%
The \np\ effects may significantly alter the patterns of CP asymmetries
in decays that are dominated by penguins in the \SM\ \GrWo.}.
We explained that in this class of models, the
unitarity triangle can be constructed model independently and the
\np\ contribution to the mixing can be disentangled from the
Standard Model one.

However, the combination of hadronic uncertainties and discrete
ambiguities puts serious obstacles in carrying out this calculation.
In particular, there is an eightfold ambiguity in the construction of
the triangle. 
In order to
get useful results, it will be necessary to reduce this ambiguity.

One way to eliminate some of the allowed solutions
can be provided by a rough knowledge of
$\cos(2\alpha-2\td)$, $\cos(2\beta+2\td)$ or $\cos2\gamma$
\ref\GrQu{Y. Grossman and H.R. Quinn, SLAC-PUB-7454, in preparation.}.
For example, $\cos(2\alpha-2\td)$ can be determined
from the CP asymmetry in $B\ra\rho\pi$
\ref\SnQu{A.E. Snyder and H.R. Quinn, Phys. Rev. D48 (1993) 2139.} and
$\cos2\gamma$ from $B\ra DK$
\ref\GrWy{M. Gronau and D. Wyler, Phys. Lett. B265 (1991) 172.}.
While a precise measurement of either of these is not expected
in the first stages of a $B$ factory, a knowledge of the sign
of the cosine is already useful for our purposes: knowing either of
sign[$\cos2(\alpha-\td)$], sign[$\cos2(\beta+\td)$] or
sign[$\cos2\gamma$] reduces the
ambiguity in $\gamma$ to fourfold. Knowing two of them reduces it to
twofold. (Knowing the three of them, however,
cannot be combined to completely eliminate the ambiguity.)

The ambiguity associated with the isospin analysis can be removed
by measuring the time dependent CP asymmetry in $B \to \pi^0\pi^0$ \GrLo.
Another way is by studying
$B\ra\rho\pi$ \refs{\SnQu,\GrQu}. Here, due to interference between
several amplitudes, isospin relations can be used to determine
$\sin 2\al$ without penguin contamination, and without any discrete
ambiguity. 

A different approach is to make further assumptions about the
\np\ that is responsible for the effects discussed above.
For example, there are many models where processes involving third
generation quarks, such as $B-\bar B$ mixing, are significantly
modified by the \np, but processes with only light quarks,
such as $K\ra\pi\nu\bar\nu$, are not. Then measurements of
$K^+\ra\pi^+\nu\bar\nu$ and $K_L\ra\pi^0\nu\bar\nu$ will provide
the true values of $R_t$ or $|\eta|$, respectively. The unitarity
triangle can be determined from these
up to a fourfold ambiguity. The additional input of $R_u$
reduces this to a twofold ambiguity.
The determination of
$\gamma$ by the methods described above will provide a {\it test} of this
class of models. It will not resolve the twofold ambiguity.

In some models
\ref\Nirxs{Y. Nir, Phys. Lett. B327 (1994) 85.}\
there is a significant contribution to both $B_d$ and $B_s$ mixing but
the ratio between the two obeys the Standard Model relation,
\eqn\BsBd{{\Delta m_{B_d}\over\Delta m_{B_s}}=F_{SU(3)}\sin^2\theta_C
R_t^2,}
where $F_{SU(3)}$ is an $SU(3)$-isospin breaking parameter.
Then, a measurement of $\Delta m_{B_s}$ will provide the correct $R_t$
and, again, the unitarity triangle can be determined, up to a twofold
discrete ambiguity, from $R_u$ and $R_t$. The determination of
$\gamma$ by our analysis is in this case, again, a test and
will not resolve the twofold ambiguity. Note, however, that in most
models where the {\it ratio} between $B_d$ and $B_s$ mixing obeys
\BsBd, the {\it phases} in the $B_s,B_d$ mixing amplitudes are
the same as in the \SM, namely $\td=0$. Then $r_d$ is the only new
parameter, and the whole analysis becomes trivial.

In a large class of models, $\epsK$ has only small
contributions from \np. If dominated by the Standard Model,
$\epsK$ implies that all angles of the unitarity triangle are in the
range $\{0,\pi\}$, and the ambiguity is reduced to fourfold.

Of course, one can combine several of these measurements and
assumptions to get a better handle on the true form of the
unitarity triangle. It is obvious however that the model independent
construction of the triangle, while possible in principle, will
pose a serious theoretical and experimental challenge.

\vskip 1 cm
\centerline{\bf Acknowledgments}
We thank Ronen Plesser and Helen Quinn for useful
discussions. Y.G. and M.P.W. are
supported by the Department of Energy under contract
DE-AC03-76SF00515. Y.N. is supported in part by the United States --
Israel Binational Science Foundation (BSF), by the Israel Science
Foundation, and by the Minerva Foundation (Munich).

\listrefs


\epsfxsize=306pt
\epsfbox{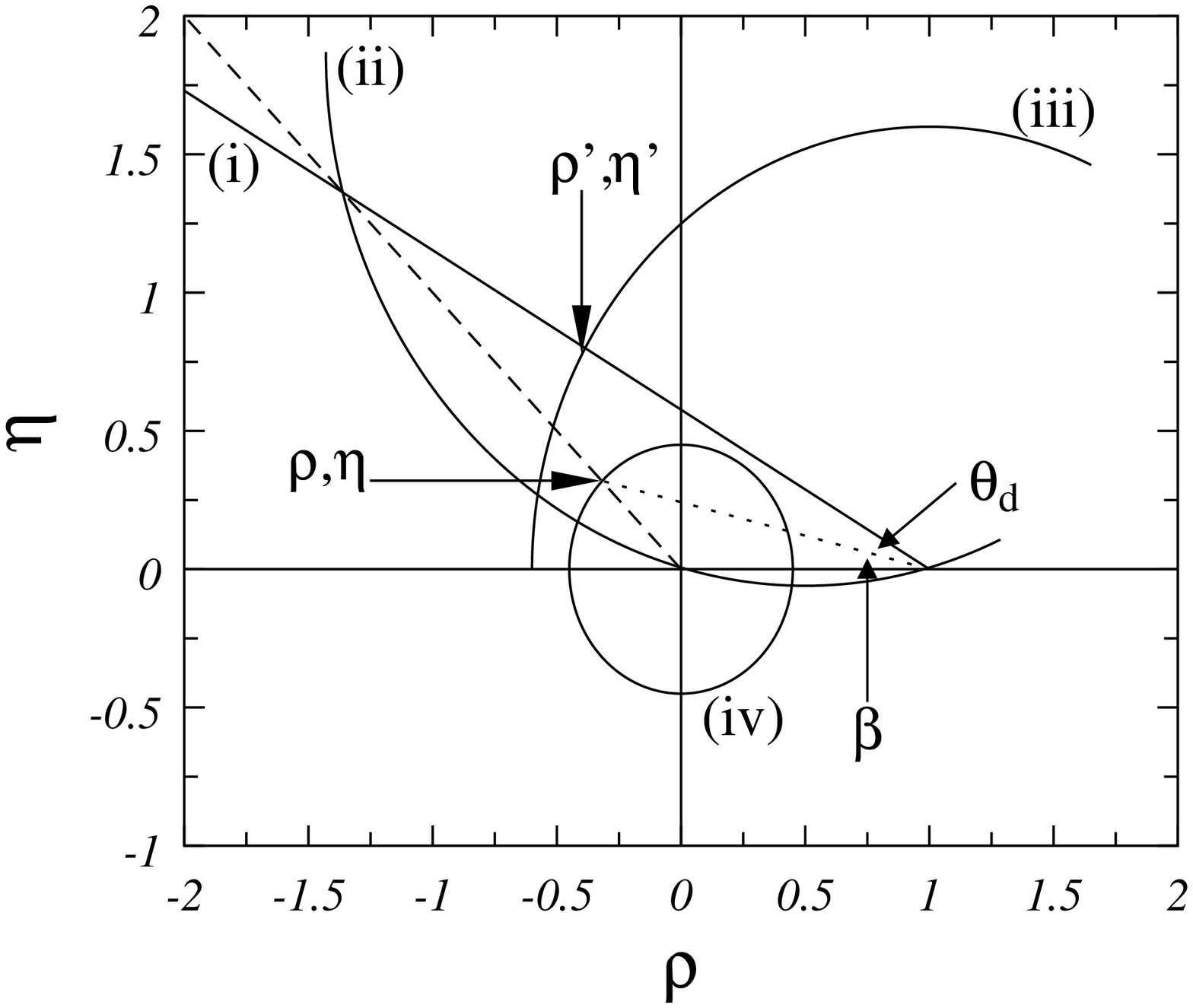}
{\rm Figure 1. }
{The model independent analysis in the $\rho-\eta$ plane:
 (i) The $\aPK$ ray; (ii) The $\aPP$ circle;
 (iii) The $x_d$ circle; (iv) The $R_u$ circle.
 The $\gamma$ ray is given by the dashed line.
 The true $\beta$ ray is given by the dotted line.
 Also shown are the true vertex of the \UT\ ($\rho,\eta$) and
 the ($\rho',\eta'$) point that serves to find $\td$ and $r_d$.}
\Date{}

\epsfxsize=306pt
\epsfbox{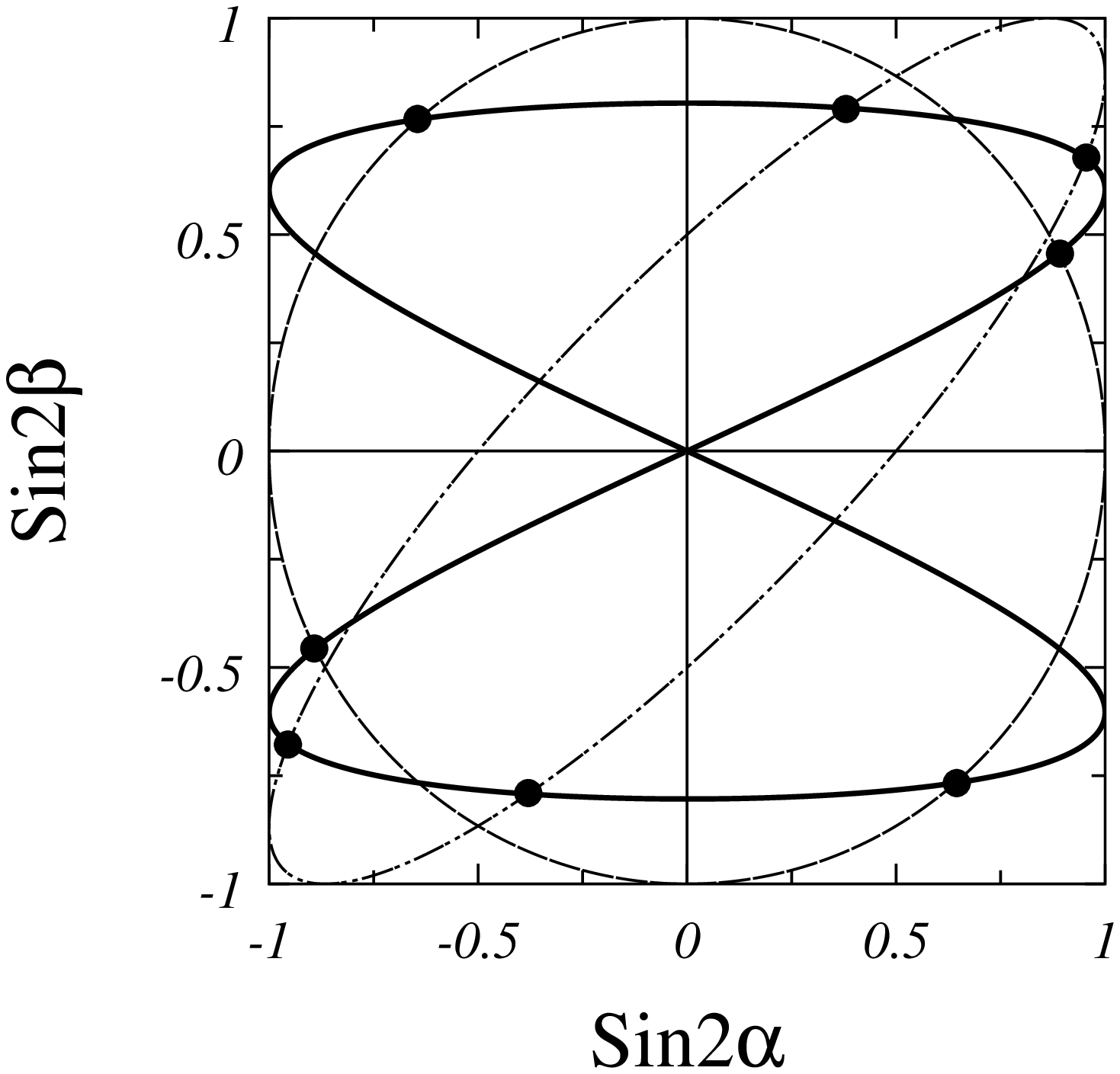}
{\rm Figure 2. }
{The $\alpha+\beta$ constraint \indsum\ and the $R_u$ constraint \Rbab\
  in the $\sin 2\alpha-\sin 2\beta$ plane. The eight possible solutions
  for the \UT\ are given by the filled circles.}

\end